**[Article Full Title]**
Study of linear energy transfer effect on rib fracture in breast patients receiving pencil-beam-scanning proton therapy

**[Short Running Title]**
LET effect on rib fracture for breast PMPT

**[Author Names]**
Yunze Yang, PhD[1,*], Kimberly R. Gergelis, MD[2,3,*], Jiajian Shen, PhD[1], Arslan Afzal, MD[2], Trey C. Mullikin, MD[4], Robert W. Gao, MD[2], Khaled Aziz, MD[2], Dean A. Shumway MD[2], Kimberly S. Corbin, MD[2], Wei Liu, PhD[1,#], Robert W. Mutter, MD[2,5,#],

[*]Co-first authors who contribute to this paper equally

[#]Co-Corresponding author

**[Author Institutions]**
[1]Department of Radiation Oncology, Mayo Clinic, Phoenix, AZ 85054, USA

[2]Department of Radiation Oncology, Mayo Clinic, Rochester, MN 55905, USA

[3]Department of Radiation Oncology, University of Rochester School of Medicine and Dentistry, Rochester, NY 14642, USA

[4]Department of Radiation Oncology, Duke Cancer Institute, Durham, NC 27710

[5]Department of Pharmacology, Mayo Clinic, Rochester, MN 55905, USA

**[Corresponding Author Name & Email Address]**
Wei Liu, PhD, Professor of Radiation Oncology, Department of Radiation Oncology, Mayo Clinic Arizona, 5777 E. Mayo Boulevard, Phoenix, AZ 85054; e-mail: Liu.Wei@mayo.edu.

Robert W. Mutter, Associate Professor of Radiation Oncology, Department of Radiation Oncology, Associate Professor of Pharmacology, Mayo Clinic Rochester, 200 1st St. SW Rochester, MN 55905; e-mail: Mutter.Robert@mayo.edu.

**[Author Responsible for Statistical Analysis Name & Email Address]**
Wei Liu, PhD, Liu.Wei@mayo.edu

**[Conflict of Interest Statement for All Authors]**
The authors declare no conflict of interest.

**[Funding Statement]**
This research was supported in part by the following resources: Arizona Biomedical Research Commission Investigator Award (R01 type), the President's Discovery Translational Program of Mayo Clinic, the Fred C. and Katherine B. Anderson Foundation Translational Cancer Research Award, The Lawrence W. and Marilyn W. Matteson Fund for Cancer Research, The Kemper

Marley Foundation, the Mayo Clinic Research Pipeline K2R Award (RWM), and the US National Cancer Institute (grant number K12 HD065987 to RWM and P30 CA015083 to MCCCC).

**[Data Availability Statement for this Work]**

Research data are stored in an institutional repository and will be shared upon request to the corresponding author.

**[Acknowledgements]**

This research was supported in part by the following resources: Arizona Biomedical Research Commission Investigator Award (R01 type), the President's Discovery Translational Program of Mayo Clinic, the Fred C. and Katherine B. Anderson Foundation Translational Cancer Research Award, The Lawrence W. and Marilyn W. Matteson Fund for Cancer Research, The Kemper Marley Foundation, the Mayo Clinic Research Pipeline K2R Award (RWM), and the US National Cancer Institute (grant number K12 HD065987 to RWM and P30 CA015083 to MCCCC).

# ABSTRACT

**Purpose:**

To study the effect of proton linear energy transfer (LET) on rib fracture in breast cancer patients treated with pencil-beam scanning proton therapy (PBS) using a novel tool of dose-LET volume histogram (DLVH).

**Methods:**

From a prospective registry of patients treated with post-mastectomy proton therapy to the chest wall and regional lymph nodes for breast cancer between 2015 and 2020, we retrospectively identified rib fracture cases detected after completing treatment. Contemporaneously treated control patients that did not develop rib fracture were matched to patients 2:1 considering prescription dose, boost location, reconstruction status, laterality, chest wall thickness, and treatment year.

The DLVH index, $V(d, l)$, defined as volume($V$) of the structure with at least dose($d$) and LET($l$), was calculated. DLVH plots between the fracture and control group were compared. Conditional logistic regression (CLR) model was used to establish the relation of $V(d, l)$ and the observed fracture at each combination of $d$ and $l$. The $p$-value derived from CLR model shows the statistical difference between fracture patients and the matched control group. Using the 2D $p$-value map derived from CLR model, the DLVH features associated with the patient outcomes were extracted.

**Results:**

Seven rib fracture patients were identified, and fourteen matched patients were selected for the control group. The median time from the completion of proton therapy to rib fracture diagnosis was 12 months (range 5 to 14 months). Two patients had grade 2 symptomatic rib fracture while the remaining 5 were grade 1 incidentally detected on imaging. The derived *p*-value map demonstrated larger *V*(0-36Gy[RBE], 4.0-5.0 keV/µm) in patients experiencing fracture ($p<0.1$). For example, the *p* value for V(30 Gy[RBE], 4.0 keV/um) was 0.069.

**Conclusions:**

In breast cancer patients receiving PBS, a larger volume of chest wall receiving moderate dose and high LET may result in increased risk of rib fracture.

**Introduction**

Proton therapy is an emerging modality for the treatment of breast cancer due to improved heart, lung, and other normal tissue sparing compared with photon techniques[1,2]. Modern pencil beam scanning proton therapy (PBS) provides even greater conformality of dose distributions than historical aperture and compensator-based proton techniques, especially in the proximal portion of the beam path[3-5]. This greater conformality of PBS enables improved organs at risk (OAR) sparing, particularly the skin, which is attractive for breast radiotherapy planning [6,7]. Therefore, PBS is increasingly utilized for the treatment of breast cancer[1].

While cell killing with protons is primarily related to physical dose, it is also impacted by the proton linear energy transfer (LET)[8]. Protons, unlike photons, deposit most of their energy over a short distance at the end of the proton beam range, and there is evidence that the higher LET at the end of range may enhance the proton relative biological effectiveness (RBE) [8-12]. Enface or anterior oblique beams are typically used in breast cancer proton therapy planning. This beam arrangement enables planners to take advantage of the rapid energy deposition at the end of the proton range to maximize heart and lung sparing, and also reduces the sensitivity of dose conformality to respiratory motion[13-20]. With this beam distribution the ribs and intercostal muscles, which lie immediately posterior to the breast, chest wall and axillary clinical target volumes (CTVs) have potential to be exposed to dose distributions with high LET and physical dose overlap. A rib fracture rate of up to 7% (grade 1 CTCAEv4.0) has been reported following proton therapy for breast cancer, higher than typically observed in the photon literature[2]. The risk of rib fracture has previously been associated with LET[21].

Currently a fixed relative biological effectiveness (RBE) value of 1.1 is employed routinely in proton therapy which ignores the potential impact of spatial variation of LET.

However, a RBE >1.1 for adverse events (AEs) associated with higher LET within OARs has been reported for rib fracture[21], rectal bleeding[22], mandible osteoradionecrosis[23,24], brain necrosis[25-28], and late-phase pulmonary changes[29]. An improved understanding of the relationship between physical dose, LET, and AEs in proton therapy planning is greatly needed to improve treatment planning.

Dose-LET volume histogram (DLVH) is a new tool that effectively combines the effects of LET and dose in patient outcomes studies[22,30,31]. Since DLVH is based on two well-defined physical quantities (i.e., dose and LET), it strategically addresses the challenges of large uncertainties in the existing models of RBE[8]. In this study, DLVH was employed to investigate the effects of dose and LET on rib fracture risk in breast cancer patients treated with PBS-based postmastectomy radiotherapy (PMRT). DLVH-based statistical methodologies, including DLVH-index-wise fixed-effect logistic regression modeling, were employed to reveal the dose/LET patterns that are potentially associated with fracture risk.

## Methods

**Patient cohort**

This study was approved by our institution research board (IRB). From a prospective registry of post-mastectomy patients treated with PBS to the chest wall and regional lymph nodes for breast cancer between 2015 and 2020, we retrospectively identified rib fracture cases detected after workup of chest wall pain or identified incidentally on imaging. Control patients without fracture were selected to match the fracture patients in a 2:1 ratio considering prescription dose, boost location, reconstruction status, laterality, chest wall thickness, and treatment year.

**Treatment planning and contouring**

The prescription dose to the chest wall and regional lymph nodes was either 50 Gy[RBE] in 25 fractions or 40 Gy[RBE] in 15 fractions. Simultaneous integrated boost (SIB) or sequential boost to lymph node and/or chest wall targets were used based on clinical risk factors at physician discretion. All patients were treated with multi-field optimized (MFO) PBS plans using two to three fields, as previously described [32-35]. Treatment plans were generated in a commercial treatment planning system (Eclipse$^{TM}$, version 15.1, Varian medical systems, Palo Alto, CA) using robust optimization[36,37] considering setup uncertainty of ±5 mm and range uncertainty of ±3%. For the CTV the planning goals were D90% ≥ 90% (priority 1) under the worst-case scenario of the plan robustness evaluation[38,39], D95% ≥ 95% (priority 2), and D0.01 cc ≤ 110% (priority 1)[34]. Treatment plans met the institutional dose volume constraints (DVCs) including target coverage and OAR dose constraints[35]. In addition, all plans were assessed using an in-house Monte Carlo biologic dose model[40]. When clinically appropriate at physician discretion attempts were made during plan optimization to limit areas of high LET and high physical dose overlap on the ribs and intercostal muscles[32] at the most posterior extent of the CTV. Plans were delivered using Hitachi PROBEAT-V proton therapy system (Hitachi, Tokyo, Japan).

**Dose-LET volume histogram (DLVH)**

Recently, we have established the tool, DLVH, to study the associations of dose and LET with normal tissue toxicity[22]. To generate the DLVH both dose and LET for each voxel were considered. The dose and dose-averaged LET were calculated using an in-house Monte Carlo dose engine[41]. This dose engine has been implemented as a second check, optimization, and biological dose evaluation platform for our clinical practice[32,40,42]. DLVH index, $V(d, l)$, was defined as the volume $V$ (% for normalized volume or cc for absolute volume) of the chest wall structure with a dose of at least $d$ Gy[RBE] and an LET of at least $l$ keV/μm, and was calculated ($V(d, l) = V(\text{Dose}>d, \text{LET}>l)$). We repeated this calculation process for all combinations of $d$ and $l$ within the dose and LET ranges. A 3D volume surface plot was then established, which represents a joint cumulative histogram of dose and LET distribution of the structure, in which dose and LET are the two independent variables. For easy visualization, the 3D volume surface plot was then projected in the 2D dose and LET plane as multiple iso-volume lines, denoted as $DLv\%$.

**DLVH-based analysis**

Chest wall DLVHs for all patients were calculated. The dose from hypofractionation plans were converted to conventional fractionation using EQD2 dose and α/β=3. In this study, to evaluate the possible dose and LET effect upon rib fractures, a risk-associated chest wall structure was retrospectively contoured which incorporated ribs and intercostal muscles enclosed by the 50% prescription isodose lines (Figure 1).

Conditional logistic regression (CLR) has been widely used to analyze the effect from case-matched data to mitigate the impact of the confounding factors. In this study, CLR was applied to DLVH analysis. DLVH index $V(d, l)$ was employed as the independent variable and the end effect (fracture or not) as the binary dependent variable to establish the relationship. In the CLR model, we assumed that two levels of effects affected the end effect. The first one is a fixed effect to consider the difference of clinical factors among fracture patients, for example, the prescription dose or the presence of tissue expander reconstruction. These fixed effects are assumed to be the same as the case-matched control groups because the differences of clinical factors among fracture patients were matched to the control group. The second one is a random effect from the DLVH index difference, represented by $V(d, l)$, between fracture group and the case-matched control group. Therefore, mixed effects from both fixed and random factors were considered in this CLR model [48]. It can be expressed using the following equation:

$$logit(\text{fracture}) = \alpha_i + \beta V(d, l)$$

where $i$ was the number of fracture patient, $\beta$ was the regression coefficient, $\alpha_i$ was the fixed effect from the corresponding case-matched patient group (the fracture patient and the corresponding control patients) $i$. Logit was the inverse of the sigmoid function, used to establish the CLR model. The *p*-value of the CLR coefficient $\beta$ was of our interest: The lower the *p*-value is, the higher the probability of a non-zero coefficient $\beta$ is, indicating a more profound impact from the DLVH index, $V(d, l)$.

For each combination of *d* and *l*, one CLR model was established. To seek for all outcomes-related DLVH indices and gain clinical insight, all DLVH indices in the range of the dose and LET of DLVH were employed one by one to establish the CLR models. After this repeating process, a 2D *p*-value map was generated for all *p*-values related to the coefficients $\beta$ of all models. Each $p(d, l)$ represented the *p*-value of the coefficient $\beta$ from the CLR model using the DLVH index $V(d, l)$ with dose *d* and LET *l*. By looking at the *p*-value map, we were able to extract the DLVH features associated with the patient outcomes.

**Statistics**

DLVHs were calculated using Matlab 2019a (MathWorks, Inc., Natick, Massachusetts, United States). The fixed-effect logistic regression was conducted using the generated "clogit" function of R (version 4.1.2). *p*-values were obtained from the regression models and were plotted using Matlab.

**Results**

**Clinical characteristics**

We reviewed 216 patients with primary or recurrent breast cancer treated with proton PMRT, with a median follow-up of 33 months (range 1 to 68 months) and identified 7 patients who experienced rib fracture. The median time from the completion of proton therapy to rib fracture diagnosis was 12 months (range 5 to 14 months). Two had grade 2 symptomatic rib fracture while the remaining 5 were grade 1 incidentally detected on imaging.

The clinical characteristics for the 7 patients with rib fracture are displayed in Table 1. The median age at time of radiation treatment was 54 years (range 32 to 64 years). Six patients received a dose of 50 Gy[RBE] in 25 fractions to the chest wall and regional lymph nodes; of these, four received a SIB of 56.25 Gy[RBE] to the chest wall (n=2) or axillary nodes (n=2), and one received a sequential boost of 14 Gy[RBE] in 7 fractions to the chest wall. One patient was treated with 40 Gy[RBE] in 15 fractions to the chest wall and regional lymphatics on a randomized trial comparing conventional and hypofractionated PMRT [43].

As shown in Table 1, two patients had only one case matched because of unique clinical circumstances; one received a sequential boost and the other was treated for recurrent disease. One other fracture patient was matched with 4 control patients to compensate for the overall patient number. Thus, the entire cohort consisted of 7 fracture patients and 14 controls.

**Chest wall DLVH comparison between fracture patients and controls**

Figure 2a displays the 3D surface plots of DLVH from a representative fracture patient along with a matched control. In this plot, the normalized volume in the z-axis represents the integral volume from both dose (x-axis) and LET (y-axis). The volume at dose $d$ Gy and LET $l$ keV/μm represents the volume that has a dose of at least *d* Gy and LET of at least *l* keV/μm. The normalized volume thus forms a 3D surface, with unity volume at 0 Gy and 0 keV/μm. Two red lines were drawn in Figure 2a to highlight the cross section between LET of 6 keV/μm and the normalized volume. The rib fracture patient has larger volumes receiving high LET (> 6 keV/μm) than the case-matched control.

To better visualize the differences between fracture patient and the matched control, we contoured iso-volume lines (5%, 20%, 50%, 80%, 95%) of DLVH plots and projected them to the dose-LET plane (Figure 2b). This helped us to observe the volume change relative to dose/LET distributions. For example, in the facture patient (Figure 2b, *left* panel), 26.5% of the chest wall structure received at least 30 Gy and 5.0 keV/μm, a point demarcated on the DLVH between the 20% and 50% iso-volume lines by an orange cross. In contrast, only 10.8% of the chest wall received 30 Gy and 5.0 keV/μm in the matched control (*right* panel), as demonstrated by the location of the cross between the 5% and 20% iso-volume lines. Therefore, from the contoured DLVH plot, we were able to directly observe the differences in dose/LET distributions by looking at the shift of the iso-volume lines.

**Features associated with rib fractures**

The derived *p*-value map demonstrated the features associated with fractures using fixed-effect logistic regression models (Figure 3a): a larger *V*(0-36Gy[RBE], 4.0-5.0 keV/μm) (*p*<0.1). In other words, the volume receiving an intermediate physical dose and high LET from the fracture group is larger than that of the control group. For example, the *p* value of the representative DLVH index, V(30 Gy[RBE], 4.0 keV/um) as shown in Figure 3b, was 0.069. The patients with grade 2 and grade 1 rib fracture were represented by square and dot symbols, respectively.

Another statistically less significant feature (p<0.2) is the smaller *V*(34-48Gy[RBE], 0-3.2keV/μm) for the fracture group, shown as feature 2 in Figure 3a. The *p* value of the representative DLVH index, *V*(30 Gy[RBE], 4.0 keV/um), as shown in Figure 3b, was 0.132. These features indicate that the fracture group received a larger volume of intermediate physical dose and high LET than that of the control group but did not receive more physical dose (another factor that could cause fractures).

**Discussion**

In this study we utilize a novel DLVH-based fixed-effect logistic regression analysis method to identify LET-related dosimetric features that may be associated with rib fracture following proton PMRT. With these advanced methods, we found that larger volume of chest wall receiving high LET and a modest physical dose was most associated with rib fracture. Our work adds to the growing body of literature suggesting that consideration of LET is warranted during breast proton therapy planning and that LET optimization may have the potential to further reduce the risk of rib fracture[2,21,32] and other adverse events of therapy[29].

Rib fracture is a well-known adverse effect of photon and proton-based radiotherapy for breast cancer[44,45]. Some studies have suggested the possibility of increased rib fracture risk following proton therapy[2,46,47]. Massachusetts General Hospital reported their initial experience of proton beam therapy for patients with breast cancer requiring regional nodal irradiation[2]; two-thirds of patients were treated with PBS, whereas the remaining patients were treated with passively scattered proton therapy (PSPT). Of 70 included patients, 5 (7%) developed symptomatic or incidentally detected rib fracture. Verma et al. reported 2 of 91 (2%) patients experienced rib fracture after receiving proton beam therapy for regional nodal irradiation [46]. The majority (77%) of patients were treated with PSPT, whereas the remainder received PBS. University of Florida recently reported their experience with 8 of 250 patients (3.7%) experiencing symptomatic or incidental rib fracture after proton radiotherapy for breast cancer [47]. Of the included patients, 58% received PBS whereas the remainder were treated with PSPT or a combination of PBS and PSPT.

Recently, Gao et al. from Mayo Clinic Rochester reported clinical outcomes for primary breast cancer patients treated exclusively with conventionally fractionated PBS[35]. With a median follow-up of 4.1 years, only two grade 2 (CTCAEv4.0) fractures from a total of 127 postmastectomy patients were reported. Recognizing limitations of cross study comparisons, the authors raised the possibility that the lower fracture rate could be due to differences in planning techniques. For example, all patients were treated with PBS using two or three fields, whereas for the patient cohort reported by Massachusetts General Hospital, the typical PBS treatment consisted of a single en face beam[21], which may lead to increased volume of high LET on the chest wall at the end of proton beam range. In addition, efforts were made during Mayo Clinic Rochester treatment planning during latter years of this study to limit areas of overlapping high

LET and physical dose over the chest wall during treatment planning, facilitated by implementation of an institutional Monte Carlo biologic dose simulation that assumes a linear relationship between RBE and dose-averaged LET[32]. The current cohort of 7 rib fracture cases includes the two rib fractures that were part of the manuscript by Gao et al[35] treated with conventionally fractionated proton PMRT. In addition, the current study includes one patient treated with hypofractionation on a randomized phase 2 trial comparing conventionally fractionated versus hypofractionated proton PMRT[43], and four high-risk patients with inflammatory breast cancer (n=3) and recurrent breast cancer (n=1).

In our study we observed dosimetric features suggesting that fractures are primarily associated with higher LET for the groups with matched physical doses. Our results are consistent with work by Wang and colleagues that a constant RBE model with a generic factor of 1.1 may be inadequate for predicting rib fracture risk[21]. Of note, the location of fractures may not be exactly in the regions of highest dosimetric risk. For example, Bradley et al. reported three fractures in patients treated with proton therapy for breast cancer that developed outside of radiation fields[47]. Recent reports have suggested that regions of AEs evolve over time and can expand to include nearby voxels with low dose and low LET[24,25,27]. Therefore, in this study we assessed the impact of LET across the entire chest wall, instead of focusing only on the site of fracture [21].

There are several limitations to our study. Post-treatment screening for rib fracture is not routinely performed and thus the incidence of rib fracture may be underestimated. In addition, we use an institutional biologic dose model during treatment planning which could affect generalizability[32], and the parameters we identified may or may not be applicable to rib fracture risk following other indications for proton therapy such as reirradiation, whole breast, and partial breast irradiation where additional investigation is warranted[1,48]. Another limitation is that non-dosimetric patient features that may be associated with rib fracture such as bone mineral density, receipt of chemotherapy, age, and menopausal status were not analyzed[49,50]. Of note, cases and controls from our study were drawn from one of the largest known proton PMRT institutional experiences in the world. Still, our analysis was limited by the small patient cohort due to the rarity of rib fracture in our patients. Although a high LET volume effect was observed, the power was insufficient to demonstrate statistical significance or to perform a univariable analysis with other clinical characteristics. Pooling of data from multi-institutional collaborators will be needed to further refine the predictive model. Towards these ends we have initiated a multi-institutional collaboration to aggregate more fracture cases after PBS. Through these collaborations, we hope to derive more conclusive insights to further optimize breast cancer proton treatment planning.

**Conclusion**

Our study reveals that larger volume receiving high LET (i.e. >4 keV/um) and moderate dose may increase the risk of rib fractures in patients undergoing PBS PMRT. These preliminary results hold promise that DLVH can be translated into clinical practice. Integration of the derived DLVH features in treatment planning may potentially minimize the incidence rate of fracture and warrants further study.

**Figure Captions**

Figure 1 Contours of chest wall (yellow), rib fracture region (red), and 50% prescription isodose line (green). The 100% prescription dose is 56.25 Gy[RBE].

Figure 2 a) 3D surface plot and b) iso-volume contour plot of Dose-LET volume histogram (DLVH) of chest wall structure for one rib fracture patient (*left* panels) and one case-matched control (*right* panels). Iso-volume lines of DL5%, DL20%, DL50%, DL80% and DL95% were displayed in the dose-LET plane. The DLVH index, $V(d, l)$, was defined as $V$(% for normalized volume) of the structure with a dose of at least $d$ Gy[RBE] and an LET of at least $l$ keV/μm. For example, the orange cross indicates the fractional chest wall volume of at least 30 Gy[RBE] and 5 keV/μm are 26.5% and 10.8% for fracture patient and the matched control, respectively. The red lines in Figure 2(a) are to highlight the difference of normalized volumes that received LET>6 keV/μm.

Figure 3 a) *P*-value map of CLR models for all DLVH indices. Iso-*p*-value lines of 0.1, 0.2 and 0.5 were contoured in the map. The features correlated with rib fracture are a larger $V$(0-36Gy[RBE], 4.0-5.0 keV/μm) of the fracture patients ($p<0.1$) (indicated as featured region 1 in red circled number). Another statistically less significant features ($p<0.2$) are a smaller $V$(34-48Gy[RBE], 0-3.2keV/μm) of the fracture patients (indicated as featured region 2 in red circled number). b) Plots of two representative DLVH indices for the two features between fracture patients and the matched controls: DLVH index 1 (*top* panel): $V$(30 Gy[RBE], 4.0 keV/ μm), indicated as the dashed green circle in a); DLVH index 2 (*bottom* panel): $V$(40 Gy[RBE], 2.5 keV/μm), indicated as the dashed purple circle in a). The statistical *p* values are 0.069 and 0.132 for DLVH index 1 and index 2, respectively. For figure 3b, the square and dot symbols represent patients with grade 2 and 1 rib fractures, respectively. RF: rib fracture patients. Ctr: matched control patients.

Table 1: The patient and tumor characteristics for the patients identified with fractures.

| Patient | Age | Stage | Fracture after RT (months) | Grade of Toxicity | Boost[a] | Dose (Gy[RBE]) | Fx | Topology[c] | Laterality | Matched Control # |
|---|---|---|---|---|---|---|---|---|---|---|
| 1 | 39 | IIIC | 14 | 1 | SIB | 50/56.25 | 25 | UIQ | L | 2 |
| 2 | 63 | IIA | 11 | 2 | No | 50 | 25 | LOQ | R | 4 |
| 3 | 56 | IIA | 5 | 1 | No | 40.05[b] | 15 | UOQ | L | 2 |
| 4 | 32 | IIIC | 14 | 2 | SIB | 50/56.25 | 25 | UOQ | R | 2 |
| 5 | 64 | IIIC | 14 | 1 | SIB | 50/56.25 | 25 | LIQ | L | 2 |
| 6 | 33 | IIIC | 11 | 1 | Seq | 50/64 | 25/32 | UOQ | L | 1 |
| 7 | 54 | Recurrent | 12 | 1 | SIB | 50/56.25 | 25 | LOQ | R | 1 |

[a]SIB: simultaneous integrated boost
[b]40.05: The doses from this hypofractionation plan were converted to conventional fractionation using EQD2 dose and $\alpha/\beta=3$.
[c]UIQ: Upper inner quadrant; LOQ: Lower outer quadrant; UOQ: Upper outer quadrant; LIQ: Lower inner quadrant.

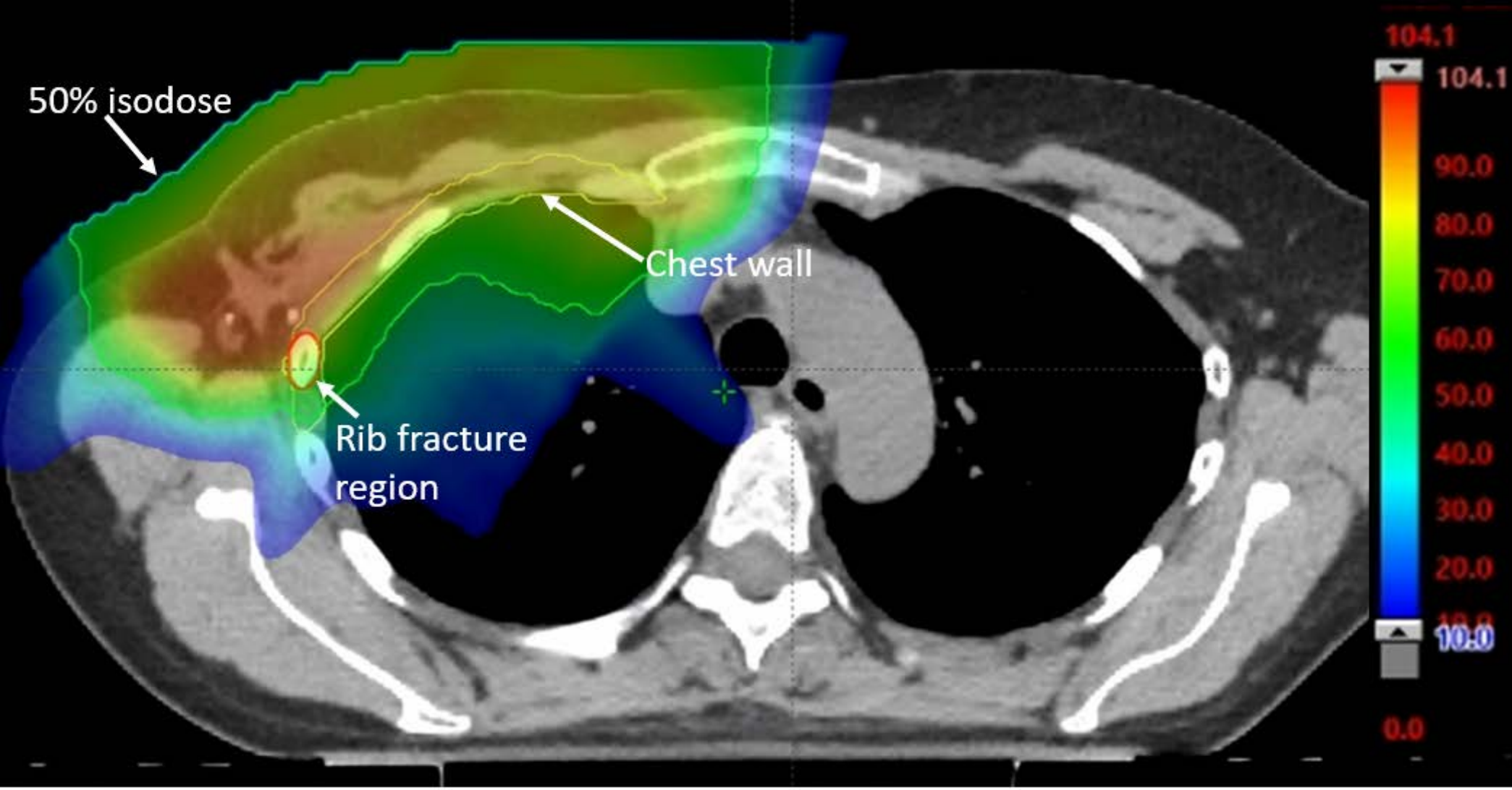

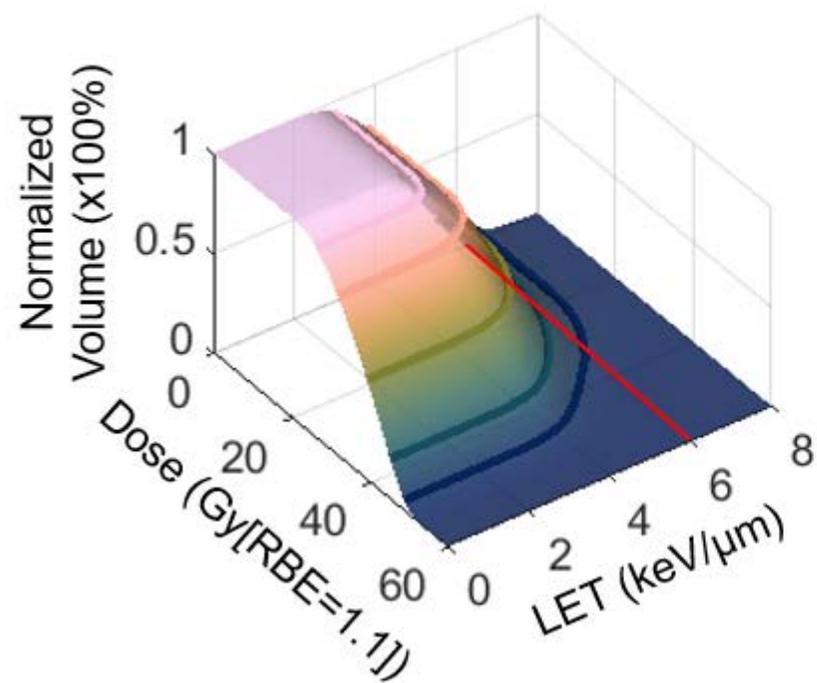 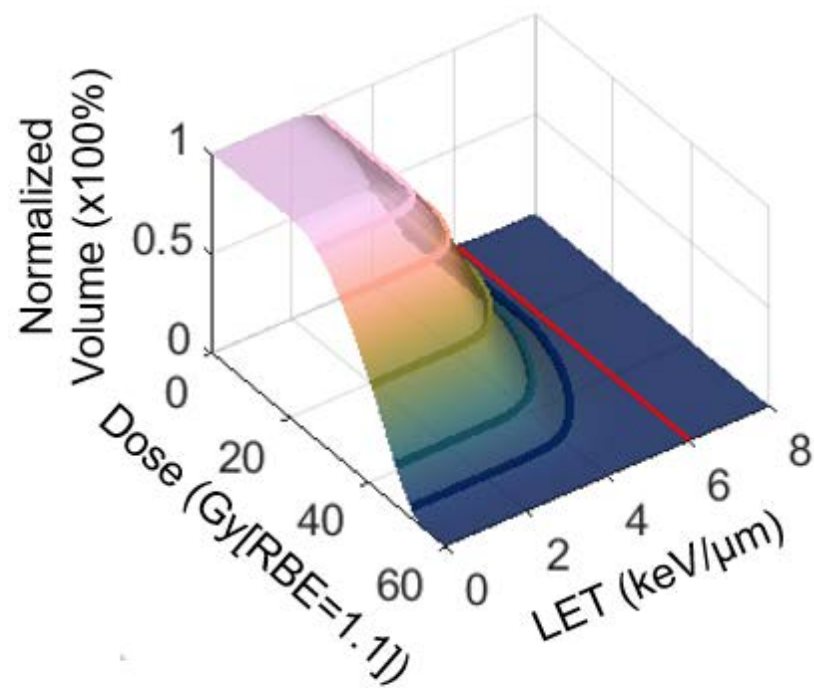
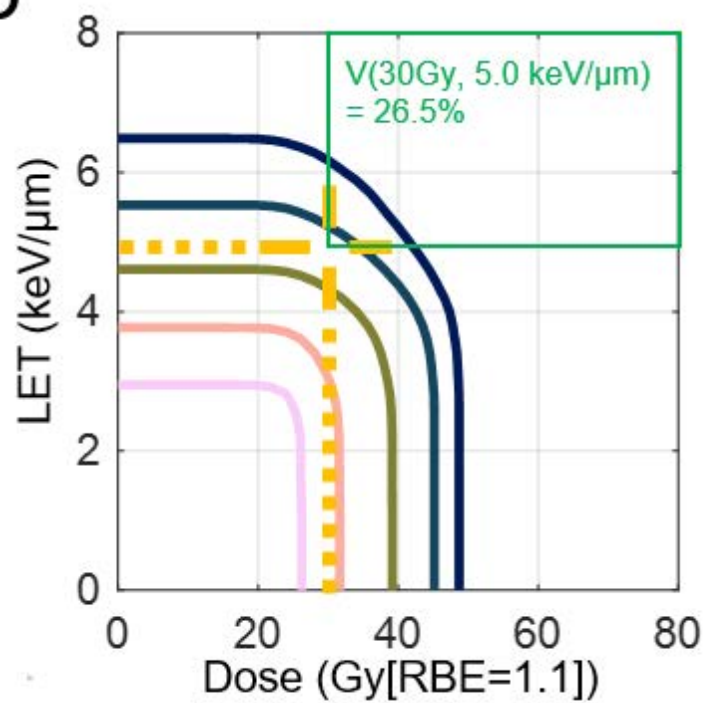 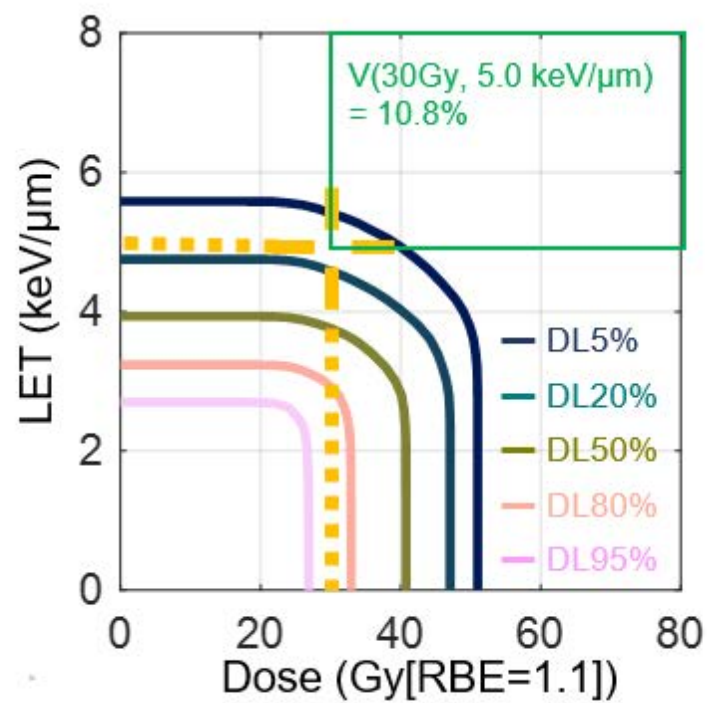

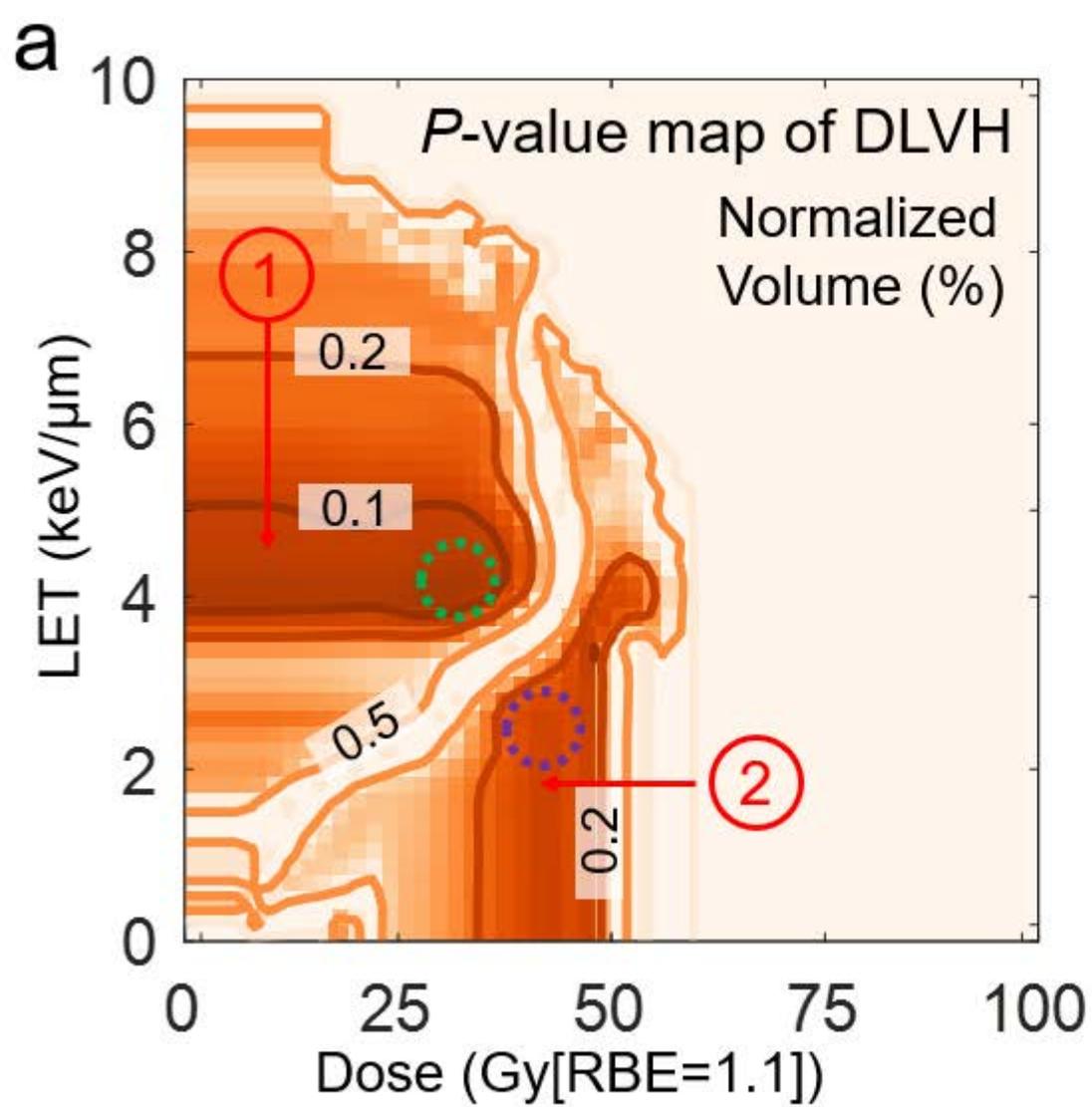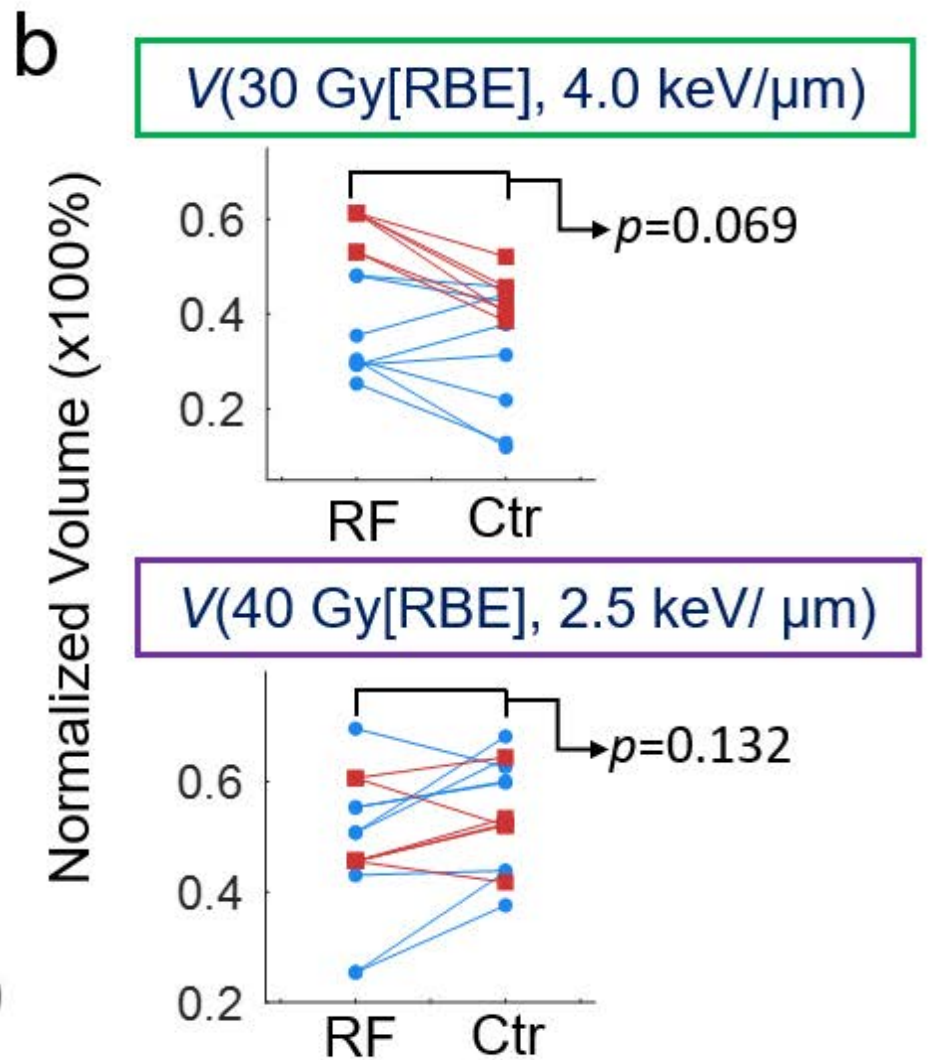